\documentclass[useAMS,usenatbib]{mn2e}
\usepackage{graphicx}
\usepackage{eufrak}
\usepackage{eucal}
\usepackage{txfonts}

\begin{document}

\title[HELGA V: the case for grain growth]{The Herschel exploitation of local galaxy Andromeda (HELGA) V: Strengthening the case for substantial interstellar grain growth}

\author[Mattsson et al.]{\parbox{\textwidth}{L. Mattsson$^{1,2}$\thanks{E-mail: mattsson@dark-cosmology.dk}, H. L. Gomez$^3$,  A. C. Andersen$^1$, M. W. L. Smith$^3$, I. De Looze$^4$,\\  M. Baes $^4$, S. Viaene$^4$, G. Gentile$^{4,5}$, J. Fritz$^4$, L. Spinoglio$^{6}$}\vspace{0.2cm}\\
$^1$Dark Cosmology Centre, Niels Bohr Institute, University of Copenhagen, Juliane Maries Vej 30, DK-2100, Copenhagen \O, Denmark\\
$^2$Nordita, KTH Royal Institute of Technology \& Stockholm University, Roslagstullsbacken 23, SE-106 91, Stockholm, Sweden\\
$^3$School of Physics \& Astronomy, Cardiff University, The Parade, Cardiff, CF24 3AA, UK\\
$^4$Department of Physics and Astronomy, University of Gent, Krijgslaan 281 - S 9, B-9000 Gent, Belgium\\
$^5$Department of Physics and Astrophysics, Vrije Universiteit Brussel, Pleinlaan 2, B-1050 Brussels, Belgium\\
$^6$Istituto di Astrofisica e Planetologia Spaziali (IAPS), Istituto Nazionale di Astrofisica (INAF), Via Fosso del Cavaliere 100, I-00133 Roma, Italy}

\date{}

\pagerange{\pageref{firstpage}--\pageref{lastpage}} \pubyear{2014}

\maketitle

\label{firstpage}

\begin{abstract} 
In this paper we consider the implications of the distributions of dust and metals in the disc of M31.  We derive mean radial dust distributions using a dust map created from {\it Herschel} images of M31 sampling the entire far-infrared peak. Modified blackbodies are fit to approximately 4000 pixels with a varying, as well as a fixed, dust emissivity index ($\beta$).  An overall metal distribution is also derived using data collected from the literature. We use a simple analytical model of the evolution of the dust in a galaxy with dust contributed by stellar sources and interstellar grain growth, and fit this model to the radial dust-to-metals distribution across the galaxy.  Our analysis shows that the dust-to-gas gradient in M31 is steeper than the metallicity gradient, suggesting interstellar dust growth is (or has been) important in M31.  We argue that M31 helps build a case for cosmic dust in galaxies being the result of substantial interstellar grain growth, while the net dust production from stars may be limited.   We note, however, that the efficiency of dust production in stars, e.g., in supernovae ejecta and/or stellar atmospheres, and grain destruction in the interstellar medium may be degenerate in our simple model. We can conclude that interstellar grain growth by accretion is likely {\it at least} as important as stellar dust production channels in building the cosmic dust component in M31.  
\end{abstract}

\begin{keywords}
ISM: clouds -- dust, extinction -- ISM: evolution -- galaxies: evolution -- galaxies: individual: M31 -- galaxies: ISM.
\end{keywords}

\section{Introduction}
The life-cycle of dust is a complex process.  It is expected that interstellar dust grains can grow by accretion in the interstellar medium  \citep[ISM; see, e.g.,][]{Ossenkopf93,Ormel09,Hirashita11} and there is observational evidence to suggest large grains are abundant in many Galactic molecular clouds \citep{Kiss06,Ridderstad06,Chapman09,Pagani10, Steinacker10}. Micrometer-sized dust grains may form in carbon-rich atmospheres of asymptotic giant branch (AGB) stars \citep{Mattsson11b}, and also in oxygen-rich AGB stars \citep{Hoefner08,Norris12}, but it is more likely that large interstellar grains have grown to such sizes inside molecular clouds since large grains produced in stars may not remain that large due to sputtering and shattering in the ISM.

In theory, shock-waves from supernovae (SNe) should destroy dust grains as these waves propagate through the ISM, but the time-scale for such dust destruction is uncertain \citep{McKee89, Draine90}. Shock destruction of dust grains is likely efficient for carbon dust, but that may not necessarily be the case for silicates \citep{Jones96,Jones04,Serra08,Jones11,Zhukovska13}.  Efficient dust destruction on short time-scales also appears inconsistent with the very high dust masses detected in high-$z$ objects \citep{Morgan03,Dwek07,Gall11,Mattsson11}.  But at least the carbon dust grains are predicted to survive in the ISM for typically not more than a few hundred Myr \citep{Jones96,Jones04,Serra08,Jones11}, which indicate a need for some kind of replenishment mechanism, rebuilding the dust component \citep{Draine90,Draine09,Mattsson13}.

It is now established that core-collapse SNe are efficient dust and molecular factories \citep{Rho09,Kamenetzky13} with large masses of cold dust detected in SN ejecta \citep[see, e.g.,][]{Morgan03b,Dunne09,Matsuura11,Gomez12,Indebetouw14} although there are significant uncertainties associated with conversion from fluxes to dust masses.  It is also unclear how much of the dust actually survives and mixes with the ISM. Theoretical results suggest 90\% of the dust produced in SNe is destroyed by the reverse shock before it reaches the ISM, depending on the interstellar gas density and the grain size distribution \citep{Bianchi07}.  Uncertain destruction rates, the lack of suitable (young, resolved) remnants, in combination with possible foreground (or background) contamination from unrelated dust clouds along the line of sight, makes it difficult to confirm whether massive-star SNe are dominant dust producers in galaxies. Thus, even if the seed grains must be produced by stars, interstellar grain growth may still be needed \citep[e.g.][]{Dunne11,Mattsson12a,Asano13}, not only as a replenishment mechanism, but also for {\em producing} the bulk of the cosmic dust {\it mass}.    Independent estimates of the efficiency of interstellar grain growth are thus still important.

The dust-to-metals ratio in a galaxy may change over time as the galaxy evolves, and this can be followed using simple analytical relations based on closed-box chemical evolution models with `instantaneous-recycling' \citep{Edmunds01,Mattsson11,Mattsson12a}, or more complex modelling \citep[e.g.,][]{Dwek98}.  Regardless of the complexity of the model, these works show that the dust abundance may not necessarily follow the metal abundances in the ISM given different sources of dust and destruction. Because of the stellar origin of both metals and dust grains, the dust-to-metals gradient along a galactic disc can therefore be regarded as a diagnostic for net dust growth or net destruction of dust in the ISM. Much of the dust mass may be the result of grain growth in the ISM and passage of shocks from SNe may lead to destruction by sputtering. If growth is dominating in the ISM the dust-to-metals gradient is negative and if destruction is dominating it will be positive \citep{Mattsson12a}. If only stars produce all the dust (as well as metals) and there is no destruction of dust in the ISM, the dust-to-metals gradient is essentially flat.  The challenge is acquiring data with enough accuracy, resolution and sensitivity to perform this kind of test. 

In \cite{Mattsson12b}, this diagnostic was used on a small sample of galaxies from the {\it Spitzer} Infrared Nearby Galaxies Survey (SINGS), where dust gradients were found to be typically steeper than the corresponding metallicity gradients, suggesting very little dust destruction and significant non-stellar dust production for most of these galaxies. The dust properties of the SINGS sample were derived from a set of ultraviolet (UV) and infrared (IR) radial profiles obtained with {\it GALEX} and {\it Spitzer} combined with optical data (SDSS {\it ugriz)} -- in total 17 different photometric bands \citep{Munoz-Mateos09,Munoz-Mateos09b}. Dust masses were obtained by fitting standard spectral energy distribution (SED) models according to \citet{Draine07} to the SEDs. However, the dust masses were derived from SEDs which lacked the long wavelength (beyond 160 $\mu$m) observational data necessary to derive accurate dust masses, with possibility of considerable errors beyond the measurement errors, i.e., a model-dependent uncertainty due to insufficient constraints from the data. Moreover, it should be noted that the information regarding the dust distribution in the galaxy discs was limited as it was derived from surface brightness profiles. But a rising trend in dust-to-gas ratio versus metallicity has also recently been found on a global scale \citep{Remy-Ruyer14}, which lends support to the interpretation of \citet{Mattsson12b}.

In the case of M31 (NGC 224; Andromeda) the situation is much improved. The launch of the European Space Agency's {\it Herschel Space Observatory}, which observes in the range $55 - 671\,\mu$m \citep{Pilbratt10} with unprecedented sensitivity and angular resolution at these wavelengths, has produced a census of galaxies as seen through their {\em dust mass}. The {\it Herschel} Exploitation of Local Galaxy Andromeda (HELGA) is a survey covering a $\sim 5.5^{\circ} \times 2.5^{\circ}$ area centred on M31 \citep[further details of the HELGA survey can be found in][]{Fritz12}.  Recently, \citet{Smith12} used the HELGA observations to investigate the distribution of dust emission in M31 on spatial scales of $\sim$140\,pc, creating maps of the dust surface density and the dust emissivity index across the disk. \citet{Draine13} have also used {\it Herschel} data \citep[][Krause et al., in preparation]{Groves12} to constrain a detailed dust model of M31. The unprecedented quality and spatial detail of the HELGA dust map makes M31 the only large spiral galaxy with a well-constrained detailed dust distribution to date.

In this paper we use the HELGA dust map of M31 in combination with oxygen abundances obtained directly or indirectly from the literature (Section \ref{data}). M31 is a good test case since the HELGA data provide extraordinary spatial resolution and detail and the metallicity of the disc can (due to its proximity) be constrained by both metallicities of stars and planetary nebulae, as well as H \textsc{ii} regions. Using a simple, well tested model of galactic dust evolution (Section \ref{modelresults}), we evaluate the importance of interstellar grain growth relative to stellar dust production in our nearest neighbour (Section  \ref{results}).

\section{Observational data and trends}
\label{data}
Here we take a closer look at the HELGA data, derived by \citet{Smith12} and evaluate the average radial trends of dust emission across M31.  We combine these results with a derivation of the metallicity gradient based on a compilation of oxygen abundance data from H {\sc ii} regions, stars and planetary nebulae to investigate the dust-to-metals gradient.

\subsection{The dust and gas data sets}
\label{datadust}

{\it Herschel} observations of M31 were taken in parallel-mode with the PACS \citep{Poglitsch10} and SPIRE \citep{Griffin10} instruments observing at 100, 160, 250, 350 and 500\,$\mu$m simultaneously.  Full details of the observing strategy and data reduction can be found in \citet{Fritz12}.  The final maps at each wavelength were created with pixel sizes of 2, 3, 6, 8 and 12\,arcsec with spatial resolution of 12.5, 13.3, 18.2, 24.5, 36.0\,arcsec full width at half-maximum for the 100, 160, 250, 350 and 500\,$\mu$m maps, respectively.  In addition to the {\it Herschel} data, the 70\,$\mu$m {\it Spitzer} MIPS map published in \citet{Gordon06} was also used as an upper limit to constrain the shorter wavelength end of the SED. 

\citet{Smith12} used this data set to create a dust surface density map of M31 by modified-blackbody fits to the FIR-SED for each pixel (using the 70\,$\mu$m data as an upper limit to the hot dust component).  Only pixels with $>5\sigma$ were used in the fits.  The flux per unit area in each pixel was modelled as
\begin{equation}
\mu_\nu = {\kappa_\nu \Sigma_{\rm d} B_\nu(T_{\rm d})\over D^2},
\end{equation}
where $B_\nu$ is the Planck distribution and the emissivity/absorptivity of the dust grains is a power law $\kappa_\nu = \kappa_0 (\nu/\nu_0)^\beta$. They assumed a value for the coefficient $\kappa_0=\kappa(350\,\mu\rm m)$ of $0.192\,\rm m^2\,kg^{-1}$, which corresponds to a typical interstellar dust composition \citep{Draine03}. Keeping $\kappa_0$ unchanged is, in principle, physically inconsistent. But we chose to do so because it yields a conservative slope of the dust profile\footnote{The value $\kappa_0=\kappa(350\,\mu{\rm m}) = 0.192\,{\rm m^2\,kg^{-1}}$ is taken from table 5 in \citet{Draine03} for a model with $\beta=2$. This makes sense for our constant $\beta = 1.8$ model, but for varying $\beta$, the extrapolation does not quite hold. In M31 we have regions with $\beta > 2.0$ (inner 5 kpc) and regions with $\beta < 2.0$ (outer regions). If we were to `correct' the dust masses here using $\kappa_0$ scaled with $\beta = 2.5$ (the most extreme inner $\beta$), $\kappa_0$ would decrease by a factor of 2 and we would get a {\it higher} dust mass.  In the outer regions, scaling $\kappa_0$ with $\beta = 1$ (most extreme outer $\beta$), $\kappa_0$ would increase by a factor of 3.5 and we would get 3.5 times {\it lower} dust masses.  Thus, the dust-to-metals ratio for the varying $\beta$ case would become even steeper.}
A distance 0.785\,Mpc was also assumed \citep{McCon05}.  In the SED fitting, \citet{Smith12} initially used a fixed value of the dust emissivity index, $\beta = 1.5$ (the slope of the long-wavelength tail of the effective absorptivity/emissivity of the dust component) across the whole galaxy, but found that with a fixed value it was impossible to adequately fit the SEDs.   $\beta$ was therefore allowed to vary across M31 [see \citet{Smith12}, fig.~7] with an estimated error in $\beta$ in any pixel of $\pm 0.31$.  It is worth noting that there is a degeneracy between $\kappa_0$ and the dust-mass density $\Sigma_{\rm d}$ in the above model. It is quite likely (if not certain) that variations in $\beta$ correspond to variations in $\kappa_0$, which means that the assumed $\kappa_0$ can put a bias on the resultant dust density $\Sigma_{\rm d}$. The decrease of $\beta$ with increasing galactocentric distance in M31 can therefore mean that we are underestimating the dust mass in its central parts. We will return to this issue later.

\cite{Smith12} also created a dust-to-gas map of M31.  The gas map was obtained by combining the atomic (H{\sc i}) and molecular (H$_2$) maps (the sum of which is adopted as the `gas mass').  The atomic hydrogen was derived from the H{\sc i} moment-zero map presented in \citet{Braun09} and the molecular hydrogen was derived from CO($J=\,$1-0) observations presented in \citet{Nieten06} made with the IRAM 30m telescope (full details are provided in \citealp{Smith12}).  Note that although the CO map requires correcting to H$_2$ [the so-called X factor, here taken to be $X_{\rm CO} = 1.9\cdot 10^{20}$~mol~cm$^{-2}$(K~km~s$^{-1}$)$^{-1}$], the correction may depend on the metallicity of the galaxy \citep[e.g.][]{Sandstrom13}. Given that the molecular gas in M31 is only 7\% of the atomic hydrogen gas, we note that this correction does not affect the conclusions of this work.

\subsection{The dust and gas distribution}
\label{dustandgas}
\begin{figure}
  \resizebox{\hsize}{!}{\includegraphics{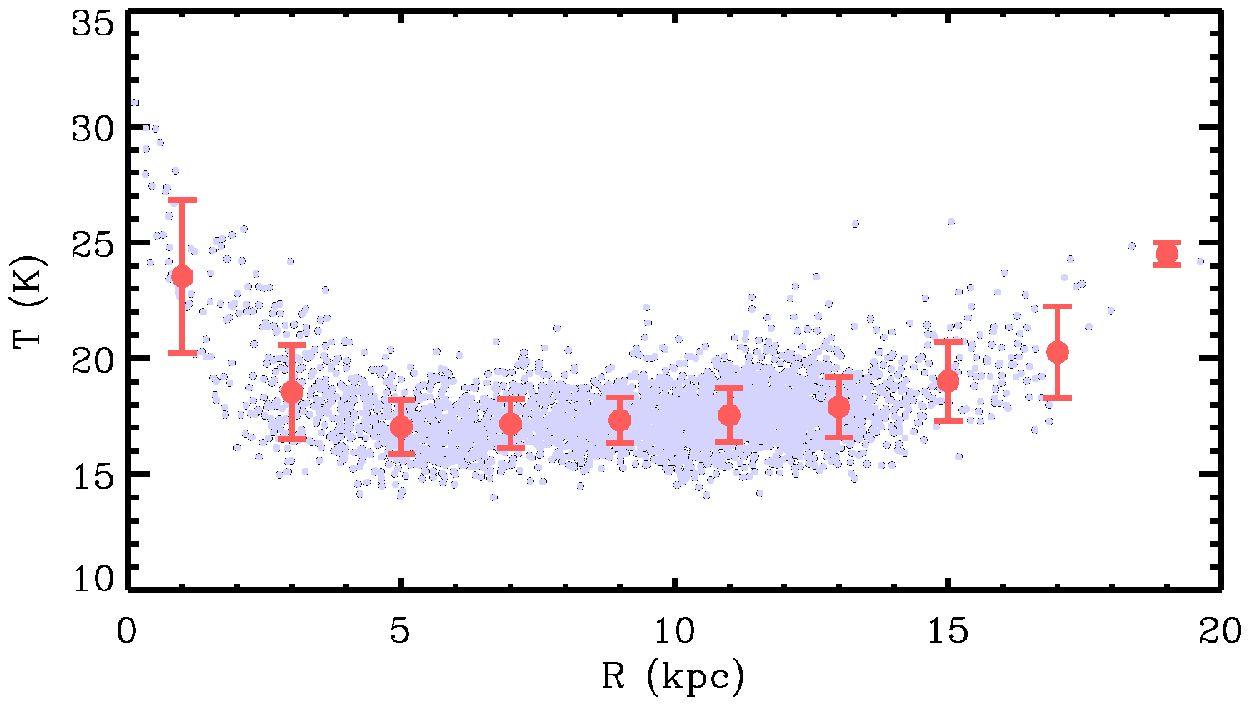}}
   \resizebox{\hsize}{!}{ \includegraphics{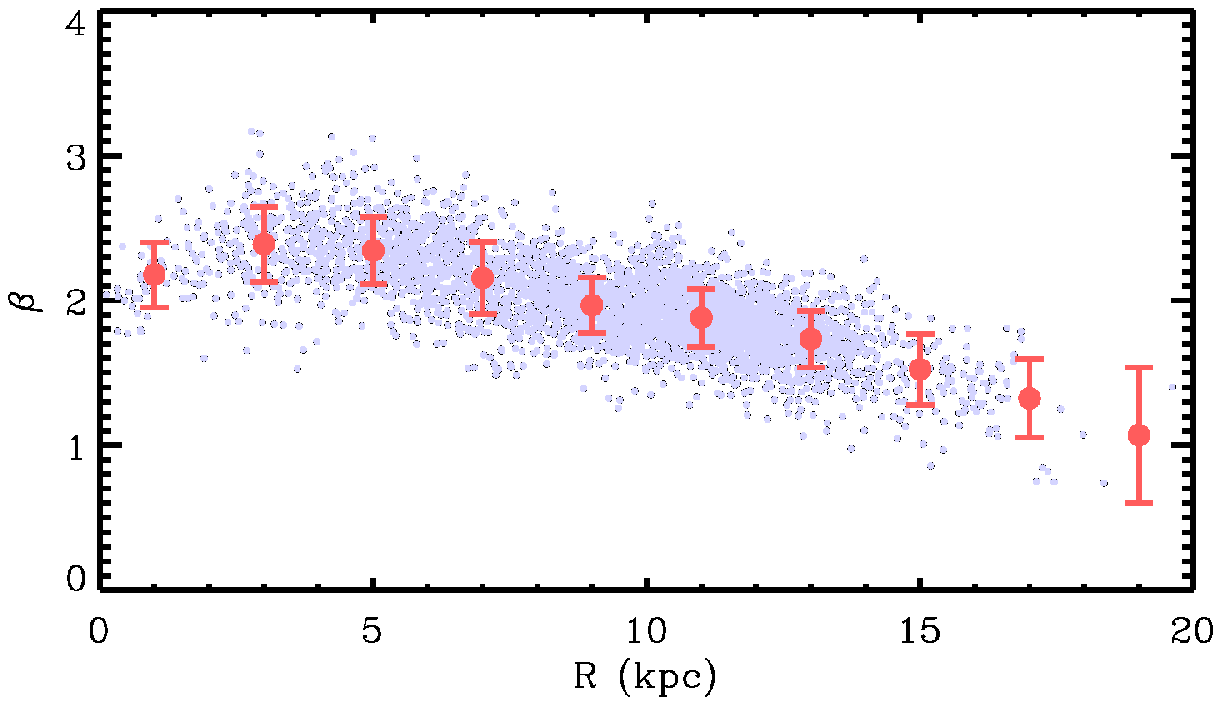}}
   \resizebox{\hsize}{!}{ \includegraphics{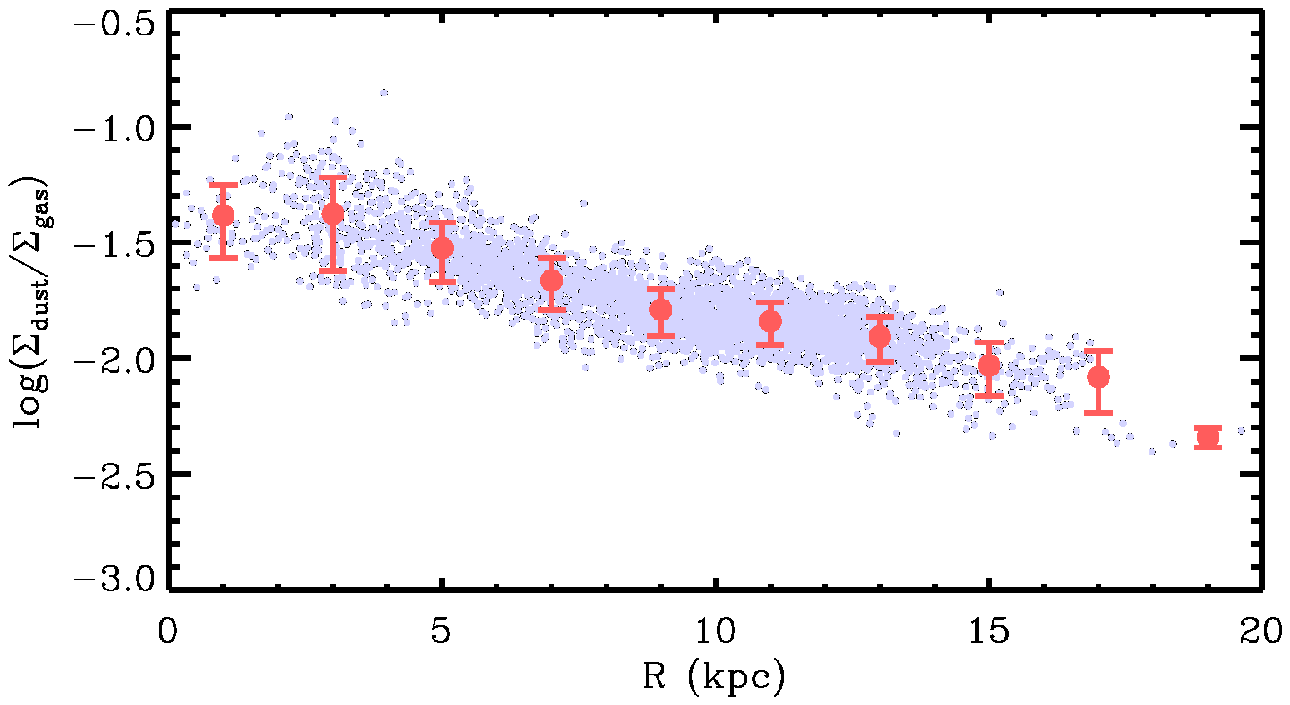}}

  \caption{\label{binned} Temperature (top panel), emissivity index $\beta$ (middle panel) and dust-to-gas ratio (bottom panel) for Andromeda as functions of galactocentric distance.  The data (grey points) are taken from the original dust map in \citet{Smith12}. 
  The red filled circles with error bars show the mean values and $1\sigma$-scatter in 2 kpc wide bins.}
  \end{figure}

The SED fitting procedure described above resulted in three parameters for each pixel across M31: the dust surface density $\Sigma_{\rm d}$, dust temperature $T_{\rm d}$ and the dust emissivity index $\beta$.   
Combining with the gas map, we also have the gas surface density in each pixel $\Sigma_{\rm gas}$.   We have binned the data in terms of consecutive radial annuli (each 2\,kpc wide, deprojected assuming an inclination of $77\deg$) and computed the mean value in each bin for all of these parameters. Based on the scatter in each radial bin, we have also computed the 1-$\sigma$ deviation from the mean values.   The resultant binned data are plotted on top of a radial projection of the dust map data in Fig. \ref{binned}. One can see in the upper panels that the dust is generally below 20\,K outside of the very centre ($R>1\,\rm kpc$), with an odd `dip' in the grain temperatures between $\sim 3-15\rm kpc$ (see top panel of Fig. \ref{binned}).  The binned $\beta$-values change significantly over the disc from $1.2$ to $2.4$ and as noted in \citet{Smith12}, increases initially out to $R \sim 3\rm \,kpc$, then decreases radially. \citet{Draine13} have also found evidence of a varying $\beta$ in M31.

It is not unexpected for the dust emissivity index ($\beta$) to vary across a galaxy, indeed this may tell us something about composition of the dust at different galactocentric distances. Low values ($\beta\sim 1$) would indicate that the dust component is dominated by amorphous carbonaceous dust \citep{Andersen99}, while higher values ($\beta\sim 2$) indicate domination by silicates or graphite \citep{Draine84}. The highest values (where $\beta > 2$) can be associated with the lowest dust temperatures, which suggests these $\beta$-values may be explained by low-temperature effects in silicates. \citet{Coupeaud11} have shown that low-temperature effects occur in the laboratory at grain temperatures below $T_{\rm d} = 12$~K, which is indeed lower than the lowest grain temperatures obtained from the SED fits, but one should bear in mind that the grain temperatures obtained from the fits represent effective temperatures for the whole dust component and not a specific dust species as in the laboratory experiments. Moreover, it also suggests that interstellar silicates are not necessarily iron rich, because silicate species such as pyroxenes ([Mg,Fe]SiO$_3$) and olivine's ([Mg,Fe]$_2$SiO$_4$) are heated more efficiently due to their higher absorptivity relative to iron-free silicates.

\begin{figure}
  \resizebox{\hsize}{!}{ \includegraphics{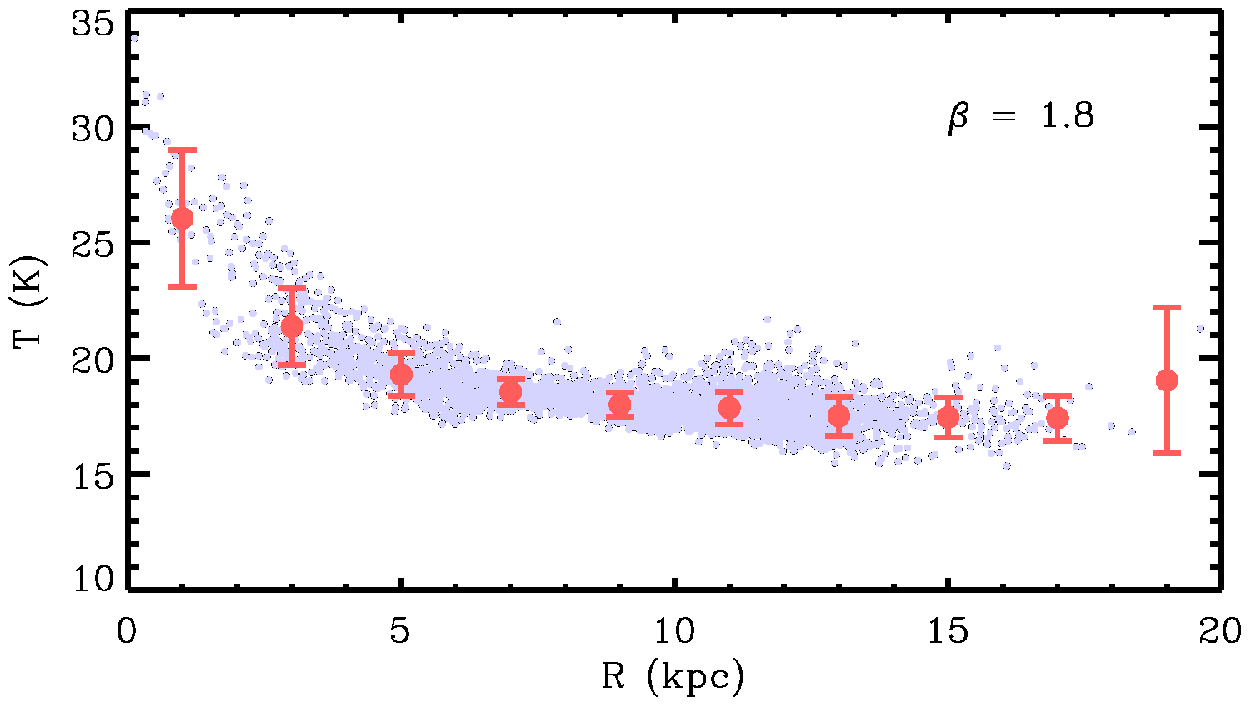}}
   \resizebox{\hsize}{!}{ \includegraphics{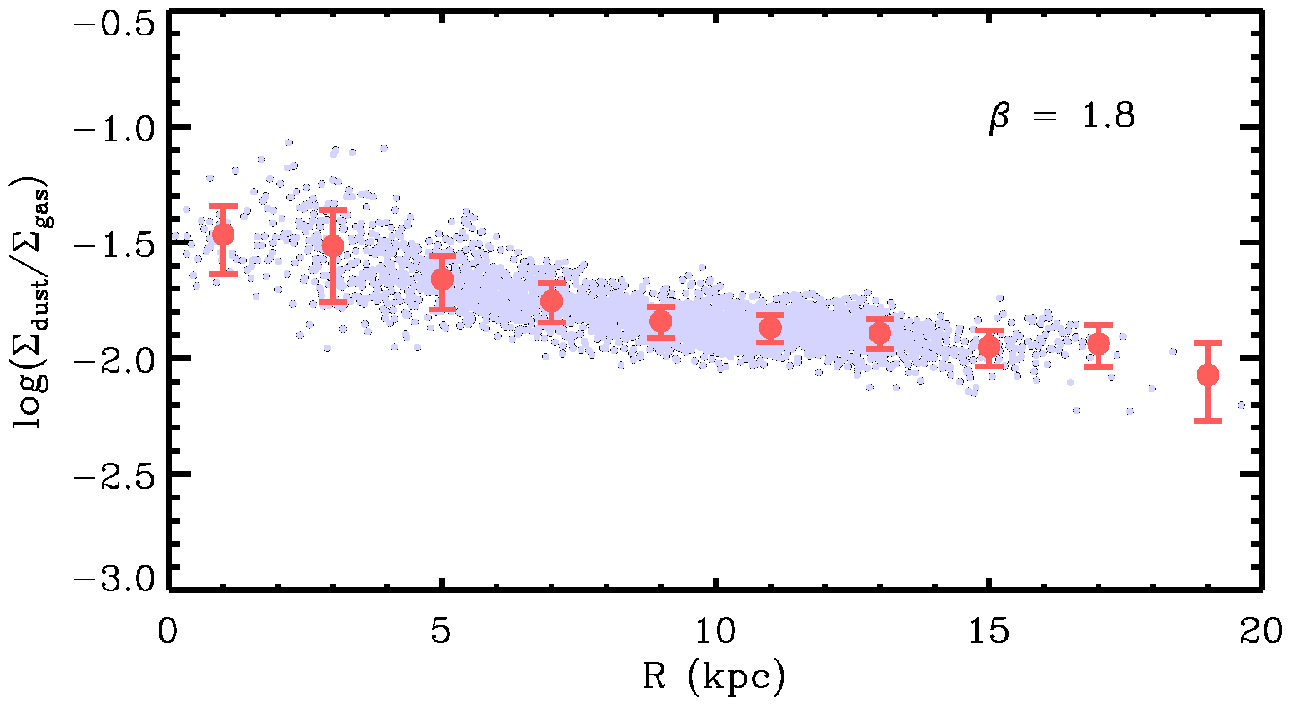}}

  \caption{\label{binned_fixedb} Same as Fig. \ref{binned}, but with properties obtained from a dust map obtained with a constant emissivity index $\beta = 1.8$. Note the difference in the dust-temperature profile compared to Fig. \ref{binned}.}
  \end{figure}

However, treating $\beta$ as a free parameter means we are at a potential risk of parameter degeneracy between the dust temperature $T_{\rm d}$ and the $\beta$-value (see discussion in Section \ref{dtgbeta}).  \citet{Smith12} demonstrated that while there is a $\beta - T_{\rm d}$ degeneracy from the fitting algorithm this does not create any systematic offsets in the value returned and therefore cannot explain the radial trends. To explore the possible effect on the dust-to-gas ratio along the disc, we have reconsidered the SED fitting to the HELGA dust map using a fixed $\beta = 1.8$. This value is a reasonable compromise, which is appropriate for the local ISM in the Galaxy \citep{Planckcoll11} and it is also in close agreement with the average $\beta$ value obtained from the varying-$\beta$ model.  The resultant dust-to-gas profile is flatter for $\beta = 1.8$ (Fig.~\ref{binned_fixedb}), though there is a clear, approximately exponential, profile in $\Sigma_{\rm d}/\Sigma_{\rm gas}$ along the disc of M31 regardless of how we treat $\beta$ (see Figs. \ref{binned} and \ref{binned_fixedb}, bottom panels). The dust temperatures are also generally higher and the temperature gradient along the disc looks more as one would expect, i.e., $T_{\rm d}$ is essentially decreasing monotonously with galactocentric distance. This means  the odd `broken' feature in the radial $T_{\rm d}$-profile is gone, with $T_{\rm d}$ simply decreasing with radius along the disc. 

        \begin{figure}
  \resizebox{\hsize}{!}{
   \includegraphics{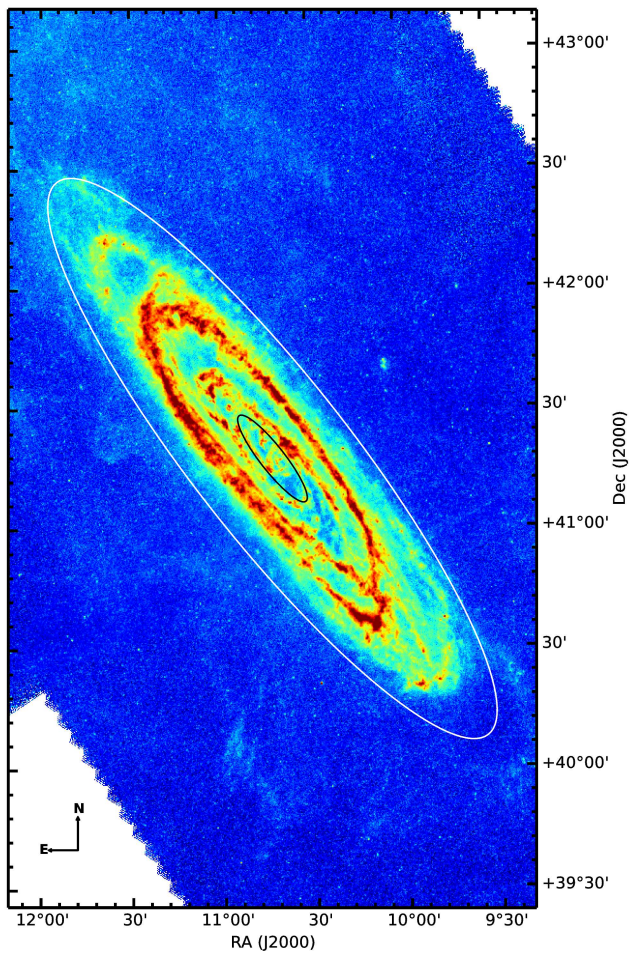}
  }
  \caption{SPIRE 250um image of M31. The white circle encloses the part of the disc for which there are meaningful dust detections. The smaller black circle marks the radius inside which \citep{Smith12} found a deviating $\beta - T_{\rm d}$ trend. \label{dustmap} }
  \end{figure}

\subsection{Metallicity gradient}
In order to estimate the overall distribution of metals in M31 we have gathered metallicity data from the literature for H {\sc ii} regions \citep{Dennefeld81,Blair82,Bresolin99,Zurita12}, young stars \citep{Venn00,Smart01,Trundle02,Lee13} and planetary nebulae \citep{Kwitter12}. We use the oxygen abundance (O/H) as a proxy for the overall metallicity. This works well for our purposes, but it is worth remembering that the conversion factor from oxygen abundance to total abundance of metals is different in low- and high-metallicity environments \citep[cf. O/Fe versus Fe/H in the Galaxy according to, e.g.,][]{Edvardsson93}. Here we adopt a universal conversion factor, which is justified by the fact that M31 has relatively similar O/H ratios (metallicity) across its disc (i.e., the conversion factor can be treated as a constant). Thus, to obtain the total metal fraction (metallicity) $Z$ we first convert the number abundances of oxygen into oxygen mass fractions using the relation $X_{\rm O} = 12\times ({\rm O/H})$ \citep{Garnett02}, in which we have implicitly assumed $M_{\rm gas} = 1.33 M_{\rm H}$. Furthermore, for M31 we may assume the oxygen typically makes up about a third of all metals \citep[which is at the low end of the possible range, see e.g.,][where 45-60\% is the suggested value]{Garnett02} and thus $Z = 3\times 12 \times ({\rm O/H})$. An oxygen fraction as low as 40\% ($Z = 2.5\times 12 \times ({\rm O/H})$) is similar to the new solar value \citep[see, e.g.,][]{Asplund09}, and an even lower fraction is expected at super-solar metallicity. One third of oxygen is therefore a reasonable assumption, which also ensures that we do not underestimate the metal content of M31 (see Section \ref{dtmgrad} for a discussion on why this is important). The adopted oxygen fraction is otherwise not critical in the present study.

Depending on the method of derivation, the derived abundance from emission spectra can vary significantly. In particular, there is a well-known offset between strong-line calibrations from empirical data and those based on photoionization models and, in general, we have to deal with the fact that oxygen abundances derived from emission spectra have no empirical {\it absolute} scale either. To obtain a homogenous set of oxygen abundances for the H {\sc ii} regions, we have re-derived O/H using the empirical strong-line calibrations by \citet{Pilyugin10} and \citet{Pilyugin11}, which are known to agree well with electron-temperature based abundances. We use the ON-calibration by \citet{Pilyugin10} for all cases where the [O~{\sc ii}]$\lambda 3727+ \lambda 3729$ line is detected with sufficient signal-to-noise (S/N~$\gtrsim 10$). In the remaining cases we use the NS-calibration by \citet{Pilyugin11}, provided the [S~{\sc ii}]$\lambda$6717+ $\lambda$6731 lines are measured. We add also a 0.1 dex correction for dust depletion, which seems to appear in H{\sc ii} regions above a certain metallicity \citep{Izotov06}\footnote {The observed phenomenon is trends in Ne/O and Ar/O versus O/H. Ne and Ar cannot be incorporated in dust, except in very small amounts as `trapped' gas inside large grains. The observed trends are therefore interpreted as dust depletion. The Ne/O trend suggests a ~0.1 dex correction at solar metallicity.}. Dust depletion cannot explain the dust-to-metals gradient, however.

\begin{figure*}
  \resizebox{\hsize}{!}{
   \includegraphics{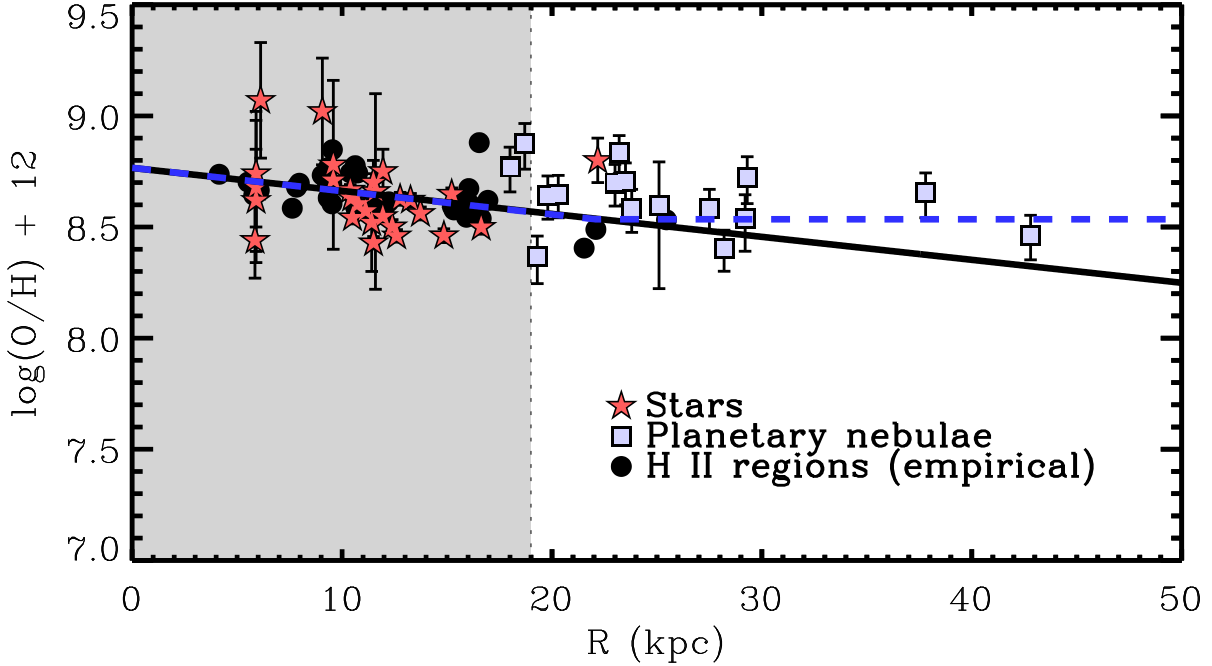}
  }
  \caption{\label{M31_OH} Oxygen abundance as a function of galactocentric distance in M31. The solid black line shows the best log-linear fit to the full set of metallicity data, while the blue dashed line show the best-fitting `broken gradient' with a constant O/H beyond $R = 23$~kpc. The grey shaded area marks the part of the disc for which there is data for both metallicity and dust-to-gas ratio from the HELGA survey.}
  \end{figure*}

In Fig. \ref{M31_OH}, we have plotted the resultant O/H ratios together with corresponding ratios derived from stars and planetary nebulae. The empirical abundances for the H {\sc ii} regions agree nicely with the stellar abundances. The abundances in H {\sc ii} regions agree with the abundances derived for planetary nebulae in the outer disc. A linear fit to all the metallicity data (see black line in Fig. \ref{M31_OH}) yields,
\begin{equation}
\label{OHgrad}
\log({\rm O/H}) + 12 = 8.77 - 0.0105\,(R/{\rm kpc}),
\end{equation}
suggesting the metallicity gradient is much flatter than the dust-to-gas gradient. The outermost data points in Fig. \ref{M31_OH} suggest a flat gradient beyond a certain galactocentric distance. A `broken gradient' fit yields almost exactly the same slope as above for $R < 23$~kpc and a flat metallicity gradient beyond $R = 23$~kpc (see the blue dashed line in Fig. \ref{M31_OH}). Since we are, in this work, only interested in the part of the disc inside a radius of 20 kpc (see the region inside the white circle in Fig. \ref{dustmap} and the grey shaded area in Fig. \ref{M31_OH}), we will in the following adopt equation (\ref{OHgrad}).
Overall, the metallicity gradient for M31 cannot be very steep regardless of the source for the metallicity data.  

\subsection{Dust-to-metals gradient}
\label{dtmgrad}
With $Z$ derived as above, we are faced with a problem \citep[which is also seen in][]{Mattsson12b}: the highest metals-to-gas ratios appears to be {\em lower} than the corresponding dust-to-gas ratios, i.e, the dust-to-metals ratio ($\zeta$) is greater than unity. This is clearly unphysical, and may arise from underestimating the metallicity, or overestimating the dust-to-gas ratio. It is not likely that we have significantly underestimated the metallicity since the errors of the abundance data are moderate and we have assumed a relatively small oxygen fraction in order to maximize the metallicity. We are thus left with an overestimated dust abundance as the only reasonable option. As mentioned in Section~\ref{datadust}, the emissivity law used in the modified blackbody fit to derive the dust surface density was anchored to the emissivity at $350\,\mu$m with $\kappa(350\,\mu{\rm m}) = 0.192\,{\rm m}^2\,{\rm kg}^{-1}$ appropriate for Milky Way-type grains. The dust composition of M31 may be somewhat different and variations in e.g., the abundance of  silicates relative to carbonaceous dust or the presence of grains with ice mantles, can easily account for an uncertainty of almost a factor of two in the dust surface density. We therefore choose to correct the derived dust density by an appropriate factor (given below, in the next paragraph) such that the dust-to-metals ratio $\zeta$ never exceeds unity.   

We suggest that the maximum dust-to-metals ratio realistically expected to be reached in the ISM is ${\rm max}(\zeta_{\rm corr}) = 0.9$, i.e., the fraction of metals in the ISM locked up in dust grains cannot exceed 90\%. The degree of dust overabundance $f$ is then defined as
\begin{equation}
f \equiv {{\rm max}(\zeta_{\rm obs}) \over {\rm max}(\zeta_{\rm corr})}  = 1.11\times {\rm max}(\zeta_{\rm obs}).
\end{equation}
The corrected dust-to-gas ratios that we will use later for our model fitting are thus $Z_{\rm d,\,corr} = Z_{\rm d}/f$. The required correction factor is $f = 2.3$ for a varying $\beta$ and $f = 1.8$ for $\beta = 1.8$. This may be interpreted as $\kappa(350\,\mu{\rm m})_{\rm corr} = f\times\kappa(350\,\mu{\rm m})$, which suggests the emissivity at  $350\,\mu$m in M31 should be $\kappa = 0.442\,{\rm m}^2\,{\rm kg}^{-1}$ in the case of a varying $\beta$ and $\kappa = 0.346\,{\rm m}^2\,{\rm kg}^{-1}$ for $\beta = 1.8$. The latter value is close to the value according to the \citet{Dunne00} model, i.e., $\kappa = 0.380\,{\rm m}^2\,{\rm kg}^{-1}$ for $\beta = 1.8$ \citep[see Eq. 5 in][]{Viaene14}. Overall, this means the dust-to-metals ratio is strongly dependent on the dust model (i.e. dust composition).

A correction of about a factor of 3 is also in agreement with the empirical emissivities obtained by \citet{Dasyra05} for three nearby spiral galaxies (NGC 891, NGC 4013 and NGC 5907).  They found that the emissivity must be roughly three times the value typically adopted for the Galaxy, which also means that Galactic sub-mm dust emissivity may be underestimated, a conclusion that is supported by our results. But note that these are edge-on spirals, where line of sight effects are largest. The elevated emissivities can certainly be disputed \citep[see, e.g.,][]{Baes10}, but it is also interesting to note that high emissivities have been found in dense, cold molecular cores, which is thought to be the result of more efficient coagulation into complex dust aggregates \citep{Stepnik03,Paradis09}.

The dust composition may of course vary along the disc, but in the present study we will, for simplicity, regard $f$ as a constant and the slope of $Z_{\rm d}$ therefore remains unchanged after the correction. The resultant dust-to-gas ratios $Z_{\rm d}$ and the dust-to-metals ratios $\zeta=Z_{\rm d}/Z$ across the disc of M31 are presented in Figs.~\ref{grad} \& \ref{zeta}.  The dust-to-gas gradient is clearly steeper than the metallicity gradient, which indicates there is a significant {\it dust-to-metals gradient} in the disc of M31. One could of course argue that $f$ may vary such that the dust-to-metals gradient flattens out and become insignificant, but there are in fact several reasons to assume the opposite, i.e., that  $f$ is likely larger in the less evolved outer regions of the disc than it is in the more evolved inner regions. Two possible reasons stand out. First, the enrichment of carbon (and thus carbonaceous dust) happens on a relatively long time-scale \citep[see, e.g., the models by][]{Carigi05, Mattsson10}.  The ratio of silicates to carbonaceous dust is therefore probably larger in the outer disc compared to the inner disc. Since carbonaceous dust grains have larger emissivity than silicate grains, $f$ would (in this scenario) {\em increase} with galactocentric distance.  Secondly, any $\beta - T_{\rm d}$ degeneracy would also lead to an overestimate of the dust mass density in the mid/outer disc more than in the inner disc because of the differences in the SED fitting results \citep[see fig. 4 in][]{Smith12}. Thus, $f$ would again {\em increase} with galactocentric distance rather than decrease, suggesting that variations in $f$ with metallicity is likely not responsible for the derived dust-to-gas slope. However, the higher $\beta$-values towards the inner disc may suggest the emissivity-law should be rescaled such that the overall emissivity is higher at small galactocentric distances. It is obviously not clear how $f$ may depend on galactocentric distance.  Assuming that the degree of dust overabundance $f$ is constant with $Z$ is therefore both a reasonable and conservative approach.

\section{Model, input and fitting}
\label{modelresults}
\begin{figure}
  \resizebox{\hsize}{!}{
   \includegraphics{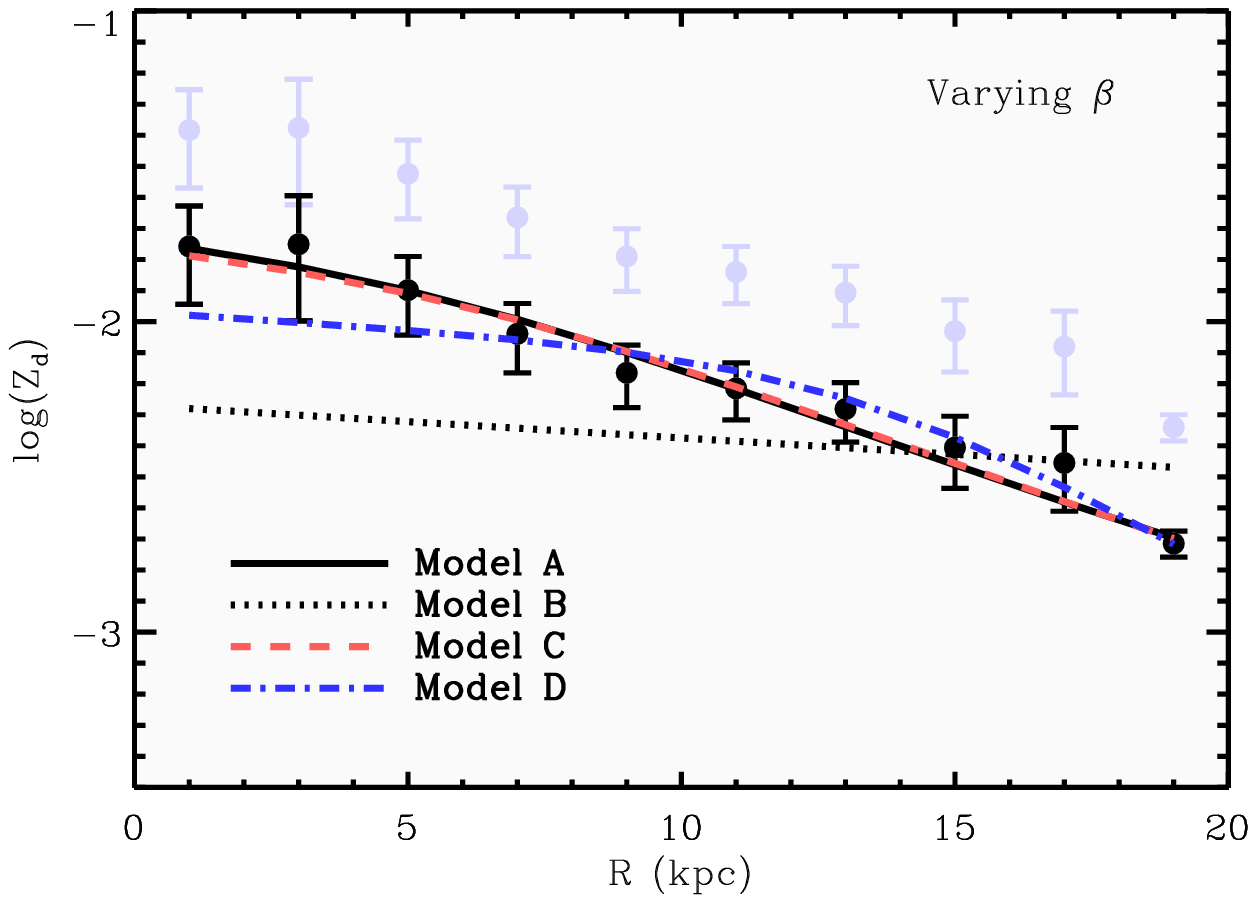}
  }
    \resizebox{\hsize}{!}{
   \includegraphics{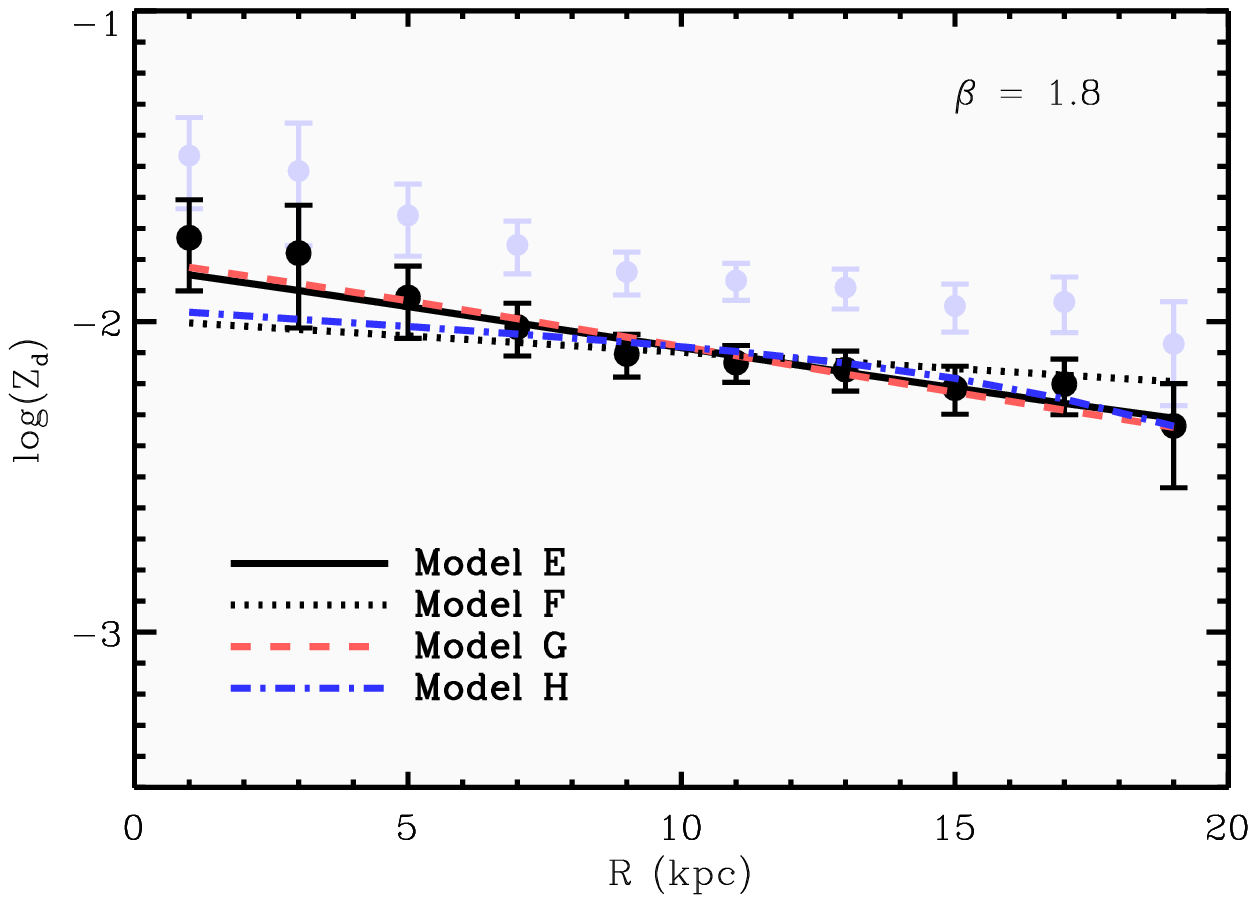}
  }
  \caption{\label{grad} Dust-to-gas ratio in M31 as function of galactocentric distance (circles) for {Top:} varying $\beta$ and {Bottom:} fixed $\beta$.   The data are compared with the best-fitting models in Table~\ref{parameters}  including stellar dust production only (dotted black lines - Models B and F) and simple models including dust growth (full drawn black, dashed red and dot-dashed blue lines - Models A, C and D).    The grey/light blue symbols in the background shows the original data before the dust-to-gas 
  ratios were corrected to account for the 
  unphysically high dust-to-metals derived (where this often exceeded unity, see Section~\ref{dtmgrad}).}
  \end{figure}

\begin{figure}
  \resizebox{\hsize}{!}{
   \includegraphics{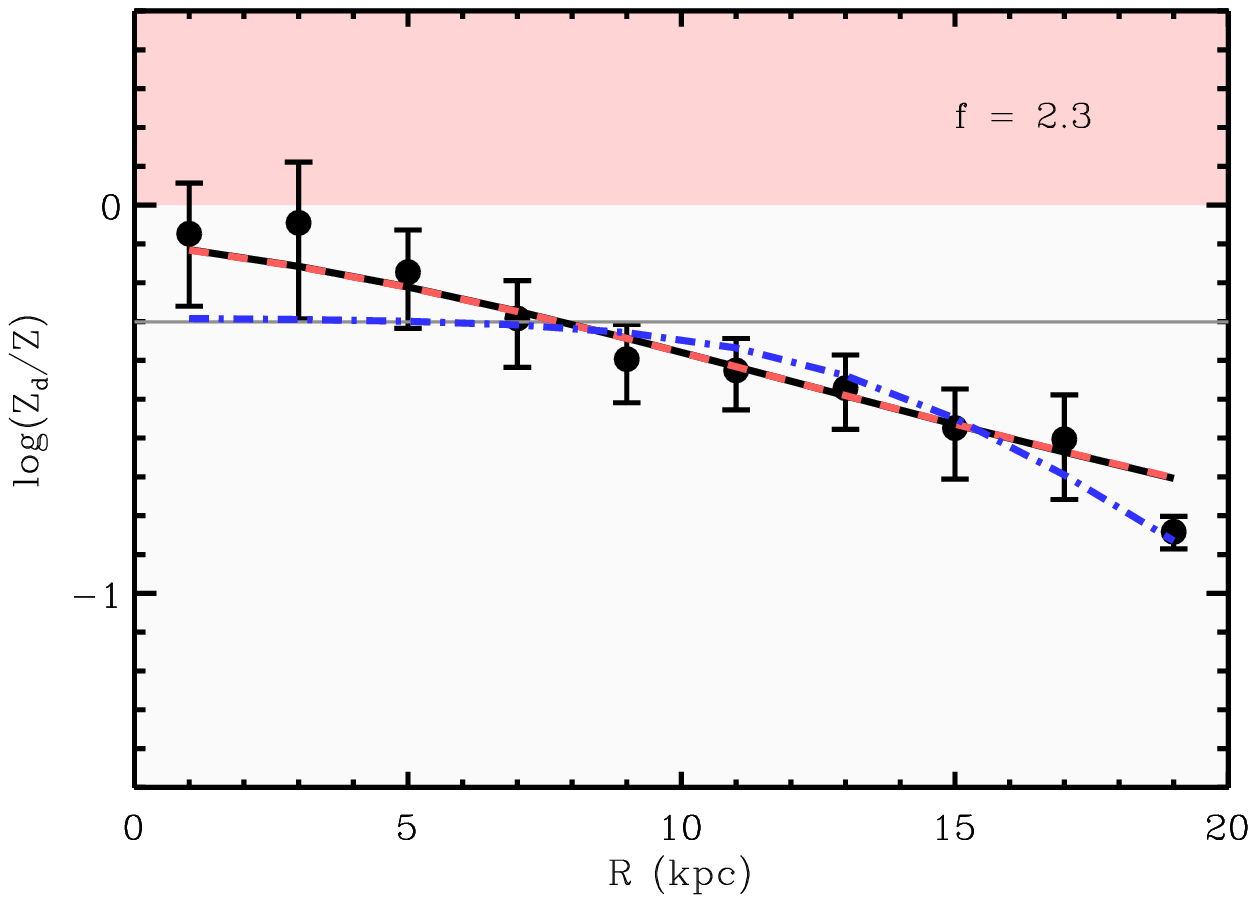}
  }
    \resizebox{\hsize}{!}{
   \includegraphics{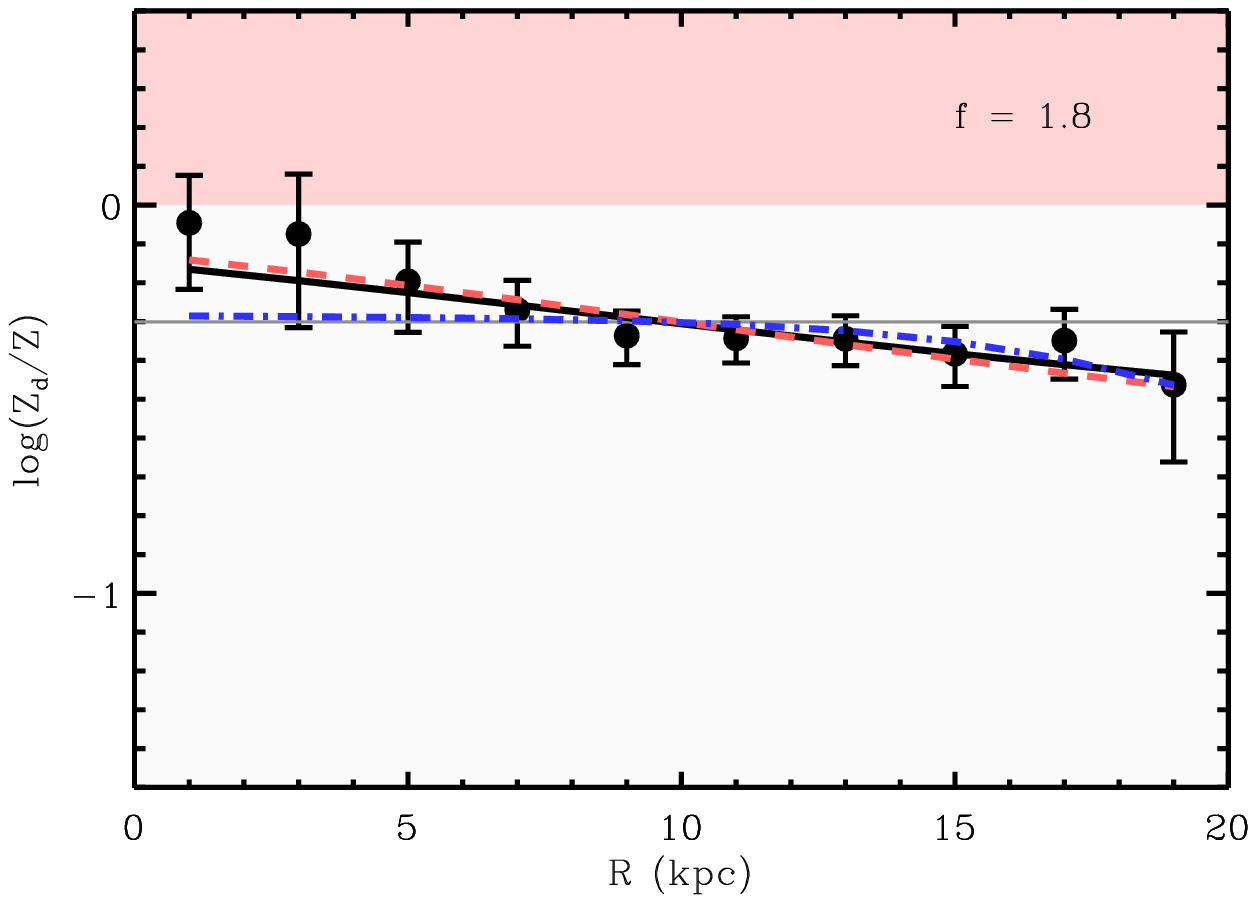}
  }
  \caption{\label{zeta} Dust-to-metals ratio in M31 as function of galactocentric distance (circles) for {Top:} varying $\beta$ and {Bottom:} fixed $\beta$.  The data are compared with the best-fitting models in Table~\ref{parameters} (as in Fig. \ref{grad}). The  light-red shaded regions correspond to dust-to-metals ratios above unity.  The thin horizontal (grey) line shows the case of an overall dust depletion of 50\%, which roughly corresponds to the dust-to-metals ratio in the Solar neighbourhood \citep{Draine07}. The `$f$-values' denote the degree of dust overabundance before correction. }
  \end{figure}

We have fitted the analytical models of dust-mass evolution derived by \citet{Mattsson12a} and \citet{Mattsson13}, which are given in terms of the so-called confluent hypergeometric Kummer-Tricomi functions \citep[][see also Appendix~\ref{theory} for further details about the models]{Kummer1837,Tricomi47} to the dust-to-gas profiles derived by \citet{Smith12}. We combine this with a metallicity profile derived from abundances as a function of galactocentric distance (Fig.~\ref{M31_OH}).  Here we briefly explain the parameters and the numerical routines for fitting the data sets in Figs.~\ref{grad} \& \ref{zeta} with the models listed in Table~\ref{parameters}.

The model has four parameters: the effective dust yield $y_{\rm d}$, the corresponding total metal yield $y_Z$, the grain-growth efficiency $\epsilon$ and the dust-destruction efficiency $\delta$. The metal yield  $y_Z$ needs to be fixed to the value obtained from the simple closed box model. The closed box model is of course not a correct model of how the metallicity in late-type galaxies evolve. But as discussed in \citet{Mattsson12a}, it is a model that works in this context since gas flows should not affect the {\it dust-to-metals ratio} very much. Using the observed O/H gradient (with a central value of O/H\,$=8.77$) derived in this work (equation~\ref{OHgrad}; Fig.~\ref{M31_OH}) and the fact that the metallicity at a galactocentric distance $R=0.4 \times R_{25}$ is known to be a good proxy for the typical metallicity of a galaxy disc \citep{Garnett02}, we have the following relationship \citep[which follows from the simple closed-box model of chemical evolution, see][]{Pagel97} for the effective metal yield,
\begin{equation}
y_Z = {Z(R=0.4\times R_{25})\over \ln(1/\mu)},
\end{equation}
where $\mu$ is the global gas mass fraction of the galaxy.  M31 is a mature spiral galaxy with relatively little gas left in the disc. Hence, we adopt a low mean gas-mass fraction of $\mu = 0.12$, which is based on various estimates found in the literature \citep[e.g.,][]{Pilyugin04,Worthey05,Tamm12}. Furthermore, we assume $R_{25} = 102.07$~arcmin \citep[see][and references therein]{Pilyugin04} such that $0.4\,R_{25} = 9.32\,\rm kpc$.  Combined with the O/H gradient, this gives $Z(R=0.4\times R_{25}) = 0.028$ and thus $y_Z = 3.5\times 10^{-3}$.  The effective stellar dust yield $y_{\rm d}$ and $\epsilon$ are treated as free parameters.

There are reasons to believe dust destruction plays a relatively minor role in the formation of a dust-to-metals gradient. This can be motivated as follows. We have seen in Section \ref{data} that there must be a significant dust-to-metals gradient in M31, which indicates significant dust growth in the ISM. M31 is also a galaxy where the metallicity is relatively similar across the disc and the dust-to-metals ratio may be close to unity in much of the inner parts (inside the white circle in Fig. \ref{dustmap}), which suggests $Z\,(1-Z_{\rm d}/Z)$ in equation. (\ref{dustz2}) is small. Thus, we expect $\epsilon \gg \delta$ since $dZ_{\rm d}/dZ$ must be positive and not too small in order for a dust-to-metals gradient to emerge. In case $\epsilon \gg \delta$, it is fair to assume a model with a negligible $\delta$, since the net effect of dust destruction would be small compared to the effect of grain growth anyway. We will therefore consider models in which $\delta = 0$ as well as where  $\delta$ is a free parameter. We also test a case where $\delta = 5.0$, which corresponds to a dust-destruction time-scale often assumed for the Galaxy \citep[$0.7-0.8$ Gyr, see][]{Jones96}.

To compare the data and the model, we used the Levenberg-Markwardt scheme for $\chi^2$-minimization. More precisely, we used the IDL-routine package MPFIT \citep[][]{Markwardt09} in combination with a numerical implementation (for IDL) of the Kummer-Tricomi functions \citep[see][]{Mattsson12a}. The initial parameter setting is in all cases $y_{\rm d} = 0.5\,y_Z$, $\epsilon = 500$ and, where used, $\delta = 5.0$. To avoid unphysical results the parameters are forced by our fitting routine to remain non-negative numbers.  In each step the fitting routine has to call a subroutine to compute the Kummer-Tricomi function $M$ for the considered argument and parameters. This may slow down the fitting considerably and even turn into a cumbersome computational effort of its own. If $M$ is to be computed on its integral form with high precision for any argument and parameter values, the number of iterations may in some cases be an inhibiting factor. Therefore, we take a short-cut, in order to maintain a reasonable computation speed. The function $M$ can be defined as an infinite series, which in turn can be evaluated in terms of factorials and the so-called $\Gamma$-function \citep{Mattsson12a}. By truncating the series once a certain precision is obtained, we have a fast and sufficiently precise numerical implementation of $M$.

Table~\ref{parameters} lists a variety of different model results fit to the derived dust-to-gas and dust-to-metal profiles in M31 (see Figs.~\ref{grad} \& \ref{zeta}).

    \begin{table*}
  \begin{center}
  \caption{\label{parameters} Settings and resultant parameter values for the model fits. $y_{\rm d}$ and $y_Z$ are the dust and metallicity yields, respectively, and $\tau_{\rm gr,\, max}$, $\tau_{\rm gr,\, min}$ are the maximum and minimum values, respectively, of the grain-growth time-scale $\tau_{\rm gr}$ as defined in Appendix \ref{theory} (equation \ref{taugr}).
  All models have $y_Z = 3.5\cdot10^{-3}$.  $\epsilon$ and $\delta$ parameterize the grain growth and dust destruction efficiencies, respectively.}
  \begin{tabular}{l|lllllllllll}
  \hline
 \multicolumn{5}{l}{Model setting:} & \multicolumn{4}{l}{Resultant parameter values:}& $\tau_{\rm gr,\, max}$& $\tau_{\rm gr,\, min}$& red. $\chi^2$\\
      & ($\beta$) & ($\epsilon$) & ($\delta$) & ~~~  & ($y_{\rm d}$) & ($y_{\rm d}/y_Z$) & ($\epsilon $& $\delta$) & (Gyr) & (Gyr)& \\
        \hline
 (A)  & Var. & Free & 0       & ~~~  & $3.06\times10^{-4}$ & $3.83\times 10^{-2}$ & 218 & - & 1.86 & 0.230 & 0.297\\
 (B)  & Var. & 0 & 0       & ~~~  & $3.29\times 10^{-3}$ & $4.12\times 10^{-1}$ & - & - & - & - & 8.92\\
 (C)  & Var. & Free & Free & ~~~  & $3.10\times 10^{-4}$ & $3.88\times 10^{-2}$ & 217 & $1.54\times 10^{-8}$ & 1.87 & 0.231 & 0.334\\
 (D)  & Var. & Free & 5.0 & ~~~  & $7.56\times 10^{-6}$ & $9.72\times 10^{-4}$ & 563 & 5.0 & 0.721 & 0.0891 & 0.745\\[1mm]
 
 (E)  & 1.8 & Free & 0       & ~~~  & $1.44\times 10^{-3}$ & $1.80\times 10^{-1}$ & 123 & - & 3.30 & 0.408 & 0.278\\
 (F)  & 1.8 & 0 & 0       & ~~~  & $3.82\times 10^{-3}$ & $4.79\times 10^{-1}$ & - & - & - & - & 0.929\\
 (G)  & 1.8 & Free & Free & ~~~  & $1.10\times 10^{-3}$ & $1..38\times 10^{-1}$ & 148 & $3.25\times 10^{-6}$ & 2.74 & 0.339 & 0.357\\
 (H)  & 1.8 & Free & 5.0 & ~~~  & $3.97\times 10^{-5}$ & $4.98\times 10^{-3}$ & 579 & 5.0 & 0.701 & 0.0866 & 0.722\\
 
  \hline
  \end{tabular}
  \end{center}
  \end{table*}

\section{Results and discussion}
\label{results}
\subsection{Model-fitting results}
Reasonable model fits can be obtained for both the varying $\beta$ case and $\beta=1.8$. There is hardly any difference between the models with $\delta = 0$ and those which have $\delta$ as an additional free parameter (see Figs. \ref{grad} and \ref{zeta} and the $\chi^2$ values given in Table \ref{parameters}). The best fit to the dust-to-gas profiles based on a fixed, as well as a varying, $\beta$ is obtained for very small $\delta$ values, which suggests the net effect of interstellar dust destruction cannot be very significant. Locking $\delta$ to a certain value (e.g., $\delta = 5$) results in a poorer, but still acceptable, fit. However, we caution the reader on the uncertainty of the simplistic model we use here - conclusions about the overall efficiency of dust destruction should not be drawn from this result. The parameter values given in Table \ref{parameters} should be taken with a grain of salt also because the fit is intrinsically uncertain. We have tried a boot-strap Monte Carlo approach to estimate the `errors' of the fitting parameters, but due to various uncertainties in the observational data, the resultant probability density functions (PDFs) cannot be used as constraints (despite 10000 iterations). The PDFs are patchy and multi-modal functions which give no meaningful statistical variance. But it is clear that one can easily vary each parameter by at least a factor of 2 and still stay within the error bars of the data.

Regardless of whether we use a fixed or varying $\beta$ there is a clear dust-to-metals gradient along the disc of M31. This is indicative of significant interstellar dust growth \citep{Mattsson12a} and we do indeed obtain relatively large $\epsilon$ values from the fits (see Table \ref{parameters}). The dust-to-metals ratio along the disc of spiral galaxies is typically not constant, but M31 (with its flat metallicity distribution) seems to have a steeper gradient than most of the SINGS spirals of comparable size \citep{Mattsson12b}. The same phenomenon is seen also in global dust-to-gas ratios and at lower metallicities in the recent results by \citet{Remy-Ruyer14}.

The favoured effective stellar dust yield $y_{\rm d}$ is in all cases with $\epsilon \neq 0$ quite small. Naively, one would interpret that as stellar dust production being rather insignificant, but as mentioned above, the simplistic model we use cannot provide very precise quantitative results due to its simplicity. The low $y_{\rm d}$-values may be due to a possible degeneracy between stellar dust production and interstellar grain destruction due to SNe in star-forming regions. 

\begin{figure}
  \resizebox{\hsize}{!}{
   \includegraphics{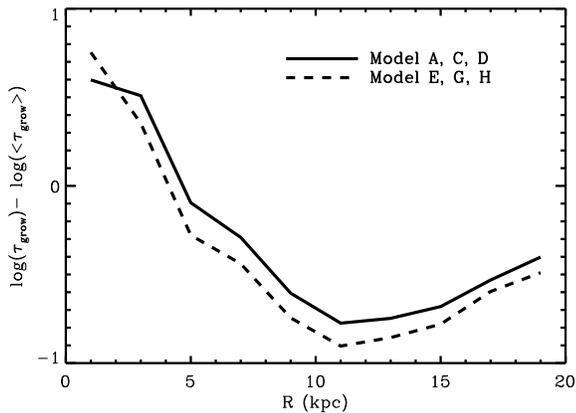}
  }
  \caption{\label{taugrplot} The grain-growth time-scale (relative to the mean value) as a function of galactocentric distance for the models where interstellar grain growth were considered. Models A, C and D (full-drawn line) correspond to the case with a free $\beta$ and models E, G, H (dashed line) correspond to $\beta = 1.8$. The time-scales are normalized to the mean value because they only differ by a constant factor, except when comparing the cases of constant and varying $\beta$, respectively.}
  \end{figure}

Since the HELGA dust map, in combination with the gas distribution, allows us to get a good handle on the Kennicutt-Schmidt (K-S) law in the M31 disc \citep{Ford13}, we have an opportunity to estimate the grain-growth time-scale $\tau_{\rm gr}$ using a consistent data set. Using the K-S law derived by \citet{Ford13} we have calculated the growth time-scale $\tau_{\rm gr}$ from equation (\ref{taugr}) in the same 2 kpc wide bins along the disc that we have used previously. The resultant $\tau_{\rm gr}$ as function of galactocentric distance is shown in Fig. \ref{taugrplot} for the cases of variable and constant emissivity index $\beta$, respectively, and the maximum and minimum values are also listed in Table \ref{parameters}. As expected, because the growth time-scale $\tau_{\rm gr}$ is anti-correlated with the gas-mass density due to its implicit dependence on the star-formation efficiency (see equations \ref{tau0} and \ref{taugr}), there is a minimum where the gas distribution has its maximum (at $R\sim 11$~kpc), but we also see a significant rise in the inner disc. These properties are found both for the case with a free emissivity index $\beta$ and for $\beta = 1.8$, i.e., the radial variation of the grain-growth time-scale (according to the models) are qualitatively the same. We note also that these grain-growth time-scales are in reasonable agreement with the results of more detailed models for the Milky Way \citep[e.g.,][who find $\tau_{\rm gr} \sim 0.5 - 1.0$ Gyr in the solar circle]{Dwek98,Zhukovska08}.

As discussed in \citet{Mattsson12a}, a dust-to-metals gradient can to some degree be the result of metallicity-dependent stellar dust production. More precisely, M-type giants (on the AGB) do not produce their own raw material for dust production, which will lead to metallicity dependence.
One could also argue that the inner regions should have more of evolved low-and intermediate-mass stars that have turned into carbon stars and thus alter the silicate-to-carbon-dust ratio, which in turn may affect the dust gradient we derive and cause an apparent metallicity dependence. However, both these issues are important only if a large fraction of the stellar dust is due to AGB stars. Since there is more and more evidence suggesting that massive stars (which produce most of the metals) are efficient dust producers, this is likely not the case. At very low metallicity there may be a threshold also for massive stars, though \citep[see the hypothesis by][]{Mattsson13}. 

Despite it is technically possible that such metallicity dependences could lead to a dust-to-metals gradient,  it can be pretty much ruled out by the result proved in Appendix A in \citet{Mattsson12a}: \\[3mm] 
{If the metallicity dependence of the effective stellar dust yield $y_{\rm d}$ is linear ($y_{\rm d}\propto Z$) and the dust-to-metals gradient is steeper than the metallicity gradient, then the slope of the dust-to-metals gradient cannot be explained by metallicity-dependent stellar dust production.}\\[3mm]
This result is, in fact, more general than so and should hold even if $y_{\rm d}$ is not linear with metallicity, and lends support to the grain-growth scenario in the present case. We note that (compare the red dashed lines in Fig. \ref{grad} with the data points in Fig. \ref{zeta}) the metallicity gradient of M31 is much flatter than its dust-to-metals gradient if the emissivity index $\beta$ is treated as a variable, while the difference is less if $\beta=1.8$. Thus, because of the flatness of the metallicity gradient in M31, there is one qualitative conclusion that may be drawn from the simple model fits we present: {the new HELGA dust map provides a good case in favour of the grain-growth scenario}.

\subsection{Caveats}
Despite the relatively firm qualitative results given above, there are a couple of caveats which we have to discuss in some detail. First, we have the problem of how to treat the emissivity index $\beta$. Is it a constant or a variable? Is it covariant with other parameters? Second, SED fitting can be done using various models of the SED. Are there one or two grain temperatures that dominate or should we consider a continuous range of grain temperatures? Below we try to address these issues.

\subsubsection{Dust-to-gas ratio: fixed versus varying $\beta$}
\label{dtgbeta}
The dust abundance data used in this paper are mainly taken from \citet{Smith12}, who choose to treat the emissivity index $\beta$ as a free parameter in their one-component SED fits, which is reasonable since $\beta$ may not be the same in every environment. However, there is a parameter degeneracy between $\beta$ and the dust temperature $T_{\rm d}$, because both affects the long-wavelength slope of the model SED \citep[see][in particular their fig. 6]{Smith12}. As one can easily see in Fig. \ref{binned} (middle panel) the mean $\beta$ values in the inner disc exceeds $\beta = 2$, which is the slope expected for silicate dust \citep{Draine84}. It is worth emphasizing what we have already mentioned in Section \ref{dustandgas}: at very low temperatures $\beta>2$ is possible \citep{Coupeaud11}, but, as discussed by \citet{Smith12}, the $\beta$--$T_{\rm d}$ relation has two branches associated with the inner and outer disc in M31. This could, in principle, be a result of the aforementioned degeneracy. If such a parameter-degeneracy problem is present, it would likely put an unphysical (and clearly unwanted) bias on the resultant dust abundances. Therefore, we have explored both fixed and varying $\beta$ in this work.

The observed SEDs are generally better fit with a varying $\beta$ and the emissivity index $\beta$ should indeed vary depending on the ratio of carbonaceous to silicate dust: the time-scale of carbon enrichment is significantly longer than that of silicon, magnesium, oxygen and other elements relevant for silicate formation \citep[see, e.g.,][]{Carigi05,Mattsson10} because it is believed that carbon is mainly produced by relatively long-lived stars that become carbon stars on the AGB. This suggests $\beta$ is lower in evolved parts of a galaxy than it is in younger parts. The $\beta$-trend with galactocentric distance should thus be increasing since the outer parts are usually less evolved than the inner parts. The fact that populations of cold grains may have large $\beta$ may work against the formation of such a trend, given that the characteristic grain temperature decreases with galactocentric distance, as one may naively assume. Adopting a constant $\beta$ over the whole disc may therefore not be completely unjustified and, more importantly, it appears to provide a lower limit to the steepness of the dust-to-gas gradient (cf. Figs. \ref{binned} and \ref{binned_fixedb}).

\subsubsection{Dust temperatures: why one-component fits are both good and bad}
Fitting a one-component modified blackbody model to the SED may not be an optimal way of estimating the dust mass. The validity of such a model depends on properties of the dust components. Sometimes there can indeed be a single dominant dust component consisting of grains of similar temperature, in which case a one-component modified blackbody model is a very good approximation. If there are two distinct dust populations with clearly different grains temperatures, such a model would be inadequate because a single component cannot capture the characteristics of a `bimodal' dust-temperature distribution. But in case the SED reflects a dust component with a continuous distribution of grain temperatures, a single component is more representative than a two-component fit, which may overestimate the contribution from the coldest grains. An overestimate of the cold component will also cause an overestimate of the dust mass, since cold grains emit significantly less radiation per unit grain mass. Also, the {\it Herschel} data provides no information about the SED beyond $500\,\mu$m, which makes it difficult to constrain the contribution from the coldest dust in a multi-temperature fit. As pointed out by \citet{Smith12}, to use a model containing dust at more than one temperature in a reliable way, we would require additional data at longer wavelengths, e.g., observations at $\sim 850\,\mu$m with SCUBA2 \citep[but see also][]{Viaene14}. 

With the above in mind, one may see the radial dust-temperature profiles in a different light: the radial dust-temperature profile, and the anomalous double-branched $\beta-T_{\rm d}$ relation in particular, does not necessarily reflect changes in the dust composition and heating sources in different parts of the disc only, but also a bias caused by assumptions about the dust temperatures underlying the SED model. The emissivity index $\beta$ is in practice just a `shape parameter' for the SED model, which is also the direct reason for the parameter degeneracy mentioned in Section \ref{dtgbeta}. In case one is trying to fit a single temperature model with a varying $\beta$ to a dust component which in reality has a continuous distribution of grain temperatures, that procedure may force $\beta$ to become smaller as the SED of a multi-temperature dust component is always wider than a single-temperature component. Obviously, this effect will also lead to an incorrect estimate of the dust mass. Hence, despite the high quality of the {\it Herschel} data, the difficulty in finding a simple but general and adequate model for the SEDs in order to obtain the dust map requires that one uses the radial dust distributions presented here \citep[and in][]{Smith12} with some caution. We note, however, that the slope of the dust-to-gas profile we obtained with a varying $\beta$, agrees well with the slope of the dust-to-gas profile in the inner disc of M31 derived by \citet{Draine13} using different data and SED fitting technique. Our result for $\beta = 1.8$, on the other hand, is in better agreement with the outer slope of the dust-to-gas profile according to \citet{Draine13}. We believe the two profiles presented here (top panels of Figs. \ref{binned} and \ref{binned_fixedb}) comprises the range of variation one may expect due to the uncertainty of the SED model.

\section{Summary and conclusions}  
We have derived and modelled distributions of dust and metals in the disc of M31 with the purpose of finding indirect evidence to hopefully distinguish between one of the two competing dust production mechanisms in late-type galaxies: stellar dust production and interstellar grain growth. The data together with simple models point towards interstellar grains growth being the most important mechanism, although firm conclusions cannot be drawn due to possible degeneracies between formation and destruction of dust associated with stars.

We have computed mean radial dust distributions from the HELGA dust map based on simple SED models with a varying as well as a fixed emissivity index $\beta$ as well as an overall metal distribution derived using data collected from the literature. In a second step, we used a simple analytical model of the evolution of the dust component in a galaxy and fit this model to the radial dust-to-gas distribution. The dust-to-gas gradient in M31 is steeper than the metallicity gradient, i.e., there is a clear negative dust-to-metals gradient along the disc, and in such cases our model suggests dust growth must  be the dominant dust-formation mechanism in the ISM of M31. Taken at face value, our model fit actually suggests the net stellar dust production is almost negligible and the build-up of the dust component in M31 should therefore be dominated by interstellar growth. 

From the above we conclude that M31 is a strong case for cosmic dust being the result of substantial interstellar grain growth, while the net stellar dust production must be limited. The efficiency of dust production in stars (most notably SNe) and the grain destruction in the ISM may be degenerate, however. Consequently, we can only conclude that interstellar grain growth by accretion is {\it at least} as important as stellar dust production channels in building the cosmic dust component in nearby galaxies. However, our result is in line with the recent (and more detailed) dust-evolution models of late-type dwarf galaxies by \citet{Zhukovska14}, which favours low condensation efficiencies in type II SNe combined with substantial grain growth in the ISM. Thus, it seems worthwhile to construct a detailed model of the radial dust distribution of M31, which is more independent of the metallicity gradient. We hope to return to this in a future publication.
 
\section*{Acknowledgements}
We thank the anonymous referee and the scientific editor (Mike Barlow) for their constructive comments which improved the final manuscript.
{\it Herschel} is an ESA space observatory with science instruments provided by European-led Principal Investigator consortia and with important participation from NASA.
Nordita is funded by the Nordic Council of Ministers, the Swedish Research Council, and the two host universities, the Royal Institute of Technology (KTH) and Stockholm University. 
The Dark Cosmology Centre is funded by the Danish National Research Foundation.
HLG is supported by the Science and Technology Facilities Council.
I. De Looze is a postdoc researcher at the FWO-Vlaanderen (Belgium).

\appendix

\section{Analytical model}
\label{theory}
In the present paper we use the analytical closed-box model described in \citet{Mattsson12a,Mattsson12b,Mattsson13}, which is based on the equation
\begin{equation}
{\partial Z_{\rm d}\over \partial Z} = {y_{\rm d}+Z_{\rm d}(\tau^{-1}_{\rm gr}-\tau^{-1}_{\rm d}) \over y_Z},
\end{equation}
where $Z$ and $Z_{\rm d}$ are the metals-to-gas (metallicity) and dust-to-gas ratios, respectively,
$\tau_{\rm d}$ is the dust destruction time scale,  $\tau_{\rm gr}$ is the interstellar dust-growth time scale and $y_{\rm d}$, $y_Z$ denote the yields \citep[dust and metals, respectively, as defined in][]{Mattsson12a}.  

According to \citet{McKee89} the dust destruction time-scale can be parameterized as
\begin{equation}
\tau_{\rm d} = {\Sigma_{\rm g}\over \langle m_{\rm ISM}\rangle\,R_{\rm SN}},
\end{equation}
where $\langle m_{\rm ISM}\rangle$ is the effective gas mass cleared of dust by each SN event, and $R_{\rm SN}$
is the SN rate. The time scale $\tau_{\rm d}$ may be approximated as \citep{Mattsson12a}
\begin{equation}
\label{taud}
\tau_{\rm d}^{-1} \approx  {\delta\over \Sigma_{\rm g}}{d\Sigma_{\rm s}\over dt},
\end{equation}
where $\delta$ will be referred to as the {dust destruction parameter, which is a measure of the efficiency of dust destruction.}  $\Sigma_{\rm s}$ and $\Sigma_{\rm g}$ denotes surface
density by mass of stars and gas, respectively.  Note that there no explicit dependence on the gas mass density $\Sigma_{\rm g}$ or the stellar mass density $\Sigma_{\rm s}$.
For a \citet{Larson98} IMF and $m_{\rm ISM} \approx 1000 M_\odot$ \citep{Jones96,Jones04}, then $\delta \approx 10$ \citep[see][]{Mattsson11} which can be regarded as an upper limit \citep{Dwek07,Gall11}.  The dust-destruction efficiency $\delta$ can also 
be calibrated to the expected dust-destruction timescale for the Galaxy, which one can assume is approximately 0.7\,Gyr \citep{Jones96}. Given that the effective Galactic gas-consumption rate is $\sim 2\rm \,M_\odot\,pc^{-2}\,Gyr^{-1}$, and the gas density is 
$\sim 8\,\rm M_\odot\,pc^{-2}$, which implies $\delta\approx 5$ \citep{Mattsson13}.

As discussed in \citet{Mattsson13}, one can modify this timescale so that the indirect effects of grain shattering are included. Small grains tend to be more easily destroyed \citep{Jones11, Slavin04} and it is therefore reasonable to assume the dust 
destruction time scale should depend on the amount of grain shattering as well. The shattering rate is to first order proportional to the square of the dust-grain density in the ISM. Thus, we may approximate the destruction time scale with the expression
\begin{equation}
\label{taud2}
\tau_{\rm d}^{-1} \approx  {\delta \over \Sigma_{\rm g}}{Z_{\rm d}\over Z_{\rm d,\,G}}{d\Sigma_{\rm s}\over dt},
\end{equation}
where $Z_{\rm d,\,G}$ is the present-day Galactic dust-to-gas ratio. 

The timescale of grain growth can thus be expressed as (see \citet{Mattsson12a}]):
\begin{equation}
\label{taugr}
\tau_{\rm gr} = \tau_0(Z) \left(1- {Z_{\rm d}\over Z}\right)^{-1}, 
\end{equation}
where, to first order, $\tau_0$ is essentially just a simple function of the metallicity and the growth rate of the stellar component, i.e.,
\begin{equation}
\label{tau0}
\tau_0 ^{-1}= {\epsilon Z\over \Sigma_{\rm g}} {d\Sigma_{\rm s}\over dt},
\end{equation}
where $\epsilon$ is a free parameter of the model.

Adopting the above scenario, with the dust-destruction timescale defined as in Eq. (\ref{taud2}), we arrive at the equation
\begin{equation}
\label{dustz2}
{dZ_{\rm d}\over dZ} = {1\over y_Z}\left\{y_{\rm d} + Z_{\rm d} \left[\epsilon \left(1-{Z_{\rm d}\over Z} \right)\,Z -\delta\,{Z_{\rm d}\over Z_{\rm d,\,G}}\right] \right\},
\end{equation}
where $y_Z$ is the metal yield. With $0\le y_{\rm d}\le y_Z$ as a basic requirement, solutions for the dust-to-gas ratio $Z_{\rm d}$ in terms of the metallicity $Z$ can
be expressed in terms of the confluent hypergeometric Kummer-Tricomi functions of the first and second kind (denoted $U$ and $M$), respectively \citep{Kummer1837,Tricomi47}. 
We refer to \citet{Mattsson12a} for further details on how such solutions are obtained. The solution to Eq. (\ref{dustz2}) can be written
\begin{equation}
\label{growthsol}
Z_{\rm d} = {y_{\rm d}\over y_Z} {
M\left[1+{1\over 2}{y_{\rm d}\over y_Z}\left(1+{1\over Z_{\rm d,\,G}}{\delta \over \epsilon}\right), {3\over 2}; {1\over 2}{\epsilon Z^2\over y_Z}\right]
\over 
M\left[{1\over 2}{y_{\rm d}\over y_Z}\left(1+{1\over Z_{\rm d,\,G}}{\delta \over \epsilon}\right), {1\over 2}; {1\over 2}{\epsilon Z^2 \over y_Z}\right]}
\,Z,
\end{equation}
where $M(a,b;z)={}_1F_1(a,b;z)$ is the Kummer-Tricomi function of the first kind, which is identical to the confluent hypergeometric function $_1F_1(a,b;z)$ . 
For comparison, we will also consider the case there is neither grain growth, nor destruction of dust in the ISM i.e. $\epsilon = \delta = 0$. We then have the trivial solution,
\begin{equation}
\label{stellarsol}
Z_{\rm d} = {y_{\rm d} \over y_Z} Z,
\end{equation}
corresponding to dust produced only by stars (e.g. from SNe and/or AGB), or a scenario where the interstellar grain growth and dust destruction are exactly balanced.

\label{lastpage}

\end{document}